\documentclass[12pt]{article}

\usepackage{scicite}
\usepackage{times}
\usepackage{graphicx,color}

\newenvironment{sciabstract}{%
\begin{quote} \bf}
{\end{quote}}

\newcounter{lastnote}

\title{Coronal transverse magnetohydrodynamic waves in a solar prominence}

\author
{
T. J. Okamoto,$^{1,2\ast}$ S. Tsuneta,$^1$
T. E. Berger,$^3$
K. Ichimoto,$^1$
Y. Katsukawa,$^1$\\
B. W. Lites,$^4$
S. Nagata,$^2$
K. Shibata,$^2$
T. Shimizu,$^5$
R. A. Shine,$^3$\\
Y. Suematsu,$^1$
T. D. Tarbell,$^3$
A. M. Title$^3$\\
\\
\normalsize{$^{1}$National Astronomical Observatory (NAOJ), Mitaka, Tokyo, 181-8588, Japan.}\\
\normalsize{$^{2}$Kwasan and Hida Observatories, Kyoto University, Yamashina, Kyoto, 607-8471, Japan.}\\
\normalsize{$^{3}$Lockheed Martin Solar and Astrophysics Laboratory,}\\
\normalsize{B/252, 3251 Hanover Street, Palo Alto, CA 94304, USA.}\\
\normalsize{$^{4}$High Altitude Observatory, National Center for Atmospheric Research,}\\
\normalsize{Post Office Box 3000, Boulder CO 80307-3000, USA.}\\
\normalsize{$^{5}$Institute of Space and Astronautical Science, Japan Aerospace Exploration Agency}\\
\normalsize{(ISAS/JAXA), Sagamihara, Kanagawa, 229-8510, Japan.}\\
\\
\normalsize{$^\ast$To whom correspondence should be addressed; E-mail:  joten.okamoto@nao.ac.jp.}\\
\\
\normalsize{{\it Recieved} 21 May 2007; {\it accepted} 06 November 2007}
}

\date{}

\begin{document} 


\baselineskip24pt


\maketitle


\begin{sciabstract}
Solar prominences are cool 10$^4$~Kelvin plasma clouds supported in the surrounding 10$^6$~Kelvin coronal plasma by as-yet undetermined mechanisms.
Observations from \emph{Hinode} show fine-scale threadlike structures oscillating in the plane of the sky with periods of several minutes. 
We suggest these transverse magnetohydrodynamic waves may represent Alfv\'en waves propagating on coronal magnetic field lines and these may play a role in heating the corona.
\end{sciabstract}


Solar prominences are classified as either quiescent or active region (AR), the latter referring to material suspended above sunspot magnetic regions.
Quiescent prominences often exist for many weeks at high solar latitudes, whereas AR prominences can be dynamic and short-lived.
They are the most enigmatic of solar structures supported by coronal magnetic field lines, sometimes erupting as the source of coronal mass ejections,
large-scale eruptions of plasma from flaring solar active regions, that can have major impacts on the terrestrial magnetic environment.
Recent ground-based observations have revealed that AR prominences have numerous small threads-like features
({\it 1}),
with continuous flow of material along the threads
({\it 2-9}).
Observations from space
({\it 10, 11})
confirm these findings and show additional dynamics related to coronal structure.


We report \emph{Hinode} Solar Optical Telescope (SOT)
({\it 12, 13})
observations of an AR prominence in a 0.3 nm broadband region centered at 396.8~nm, the H-line spectral feature of singly ionized calcium (Ca~II).
Radiation in this bandpass typically has a temperature of less than 20,000~K.


We obtained over 1 hour of continuous SOT images of NOAA AR~10921 on the west solar limb on 9 November 2006.
The images show a multithreaded AR prominence suspended above the main sunspot (Fig. 1).
Although no simultaneous H$\alpha$ images were taken,
the Ca~II H-line prominence structures are consistent with the structures seen in lower-resolution H$\alpha$ observations
({\it 14}).
The Ca~II H-line movie (movie S1) shows ubiquitous continuous horizontal motions along the prominence threads.
The origins of these flows remain unknown. Some of the flows had constant speeds of about 40 km s$^{-1}$,
whereas others accelerated monotonically or in a more complicated fashion.

The \emph{Hinode} SOT movies also reveal that many of the threads in the prominence underwent vertical
(i.e. in the plane of the sky) oscillatory motions (Fig. 2) at periods of 130 to 250~s.
The vertical oscillatory motions were coherent over lengths as long as 16,000 km.
One thread (Fig. 3A) had a vertical width of 660~km and an oscillation period of 240~s.
Comparison of the oscillation phase at various horizontal positions (Fig 3, B to F) reveals that the thread oscillated in phase along its entire length.
The vertical oscillation amplitudes of all threads we saw ranged from 400 to 1800~km, and thread widths were between 430 and 660~km (Table 1).
Because we cannot determine the angle between the plane of oscillation and the line of sight to the prominence threads,
the oscillation amplitudes as well as the horizontal velocities are minimum estimates.


The observed vertical oscillatory motion is most likely due to propagating or standing Alfv\'en waves on the horizontal magnetic field lines that compose the prominence.
An alternate hypothesis is longitudinal plasma motion along helical field lines.
Helical field line structure has been observed in many previous prominence studies 
({\it 15-17});
however, the helical field line hypothesis cannot explain the several cases we found of synchronous vertical oscillation of entire threads, for example, Fig.~3.
Alternately, if the observed threads are thin magnetic flux tubes, the observed oscillations may be fast magneto-acoustic kink modes propagating along the tubes
({\it 18, 19}).
The observations shown here lack line-of-sight Doppler velocity measurements so we cannot determine the exact oscillation mode at this point.

These field lines connect to the source regions in the photosphere 
({\it 7}),
where they are excited by the wide-spectrum p-mode oscillations originating in the convection zone.
Each field line oscillated independently, as seen in the \emph{Hinode} movies.
The synchronous oscillation along entire threads shown in Fig. 3 implies that 
we cannot distinguish the time difference at the minimum and maximum amplitudes all along the thread.
The uncertainty in phase of the oscillation is no more than 1/16 of the oscillation period,
so we estimate that the minimum wavelength of the oscillation is 16$\times$16,000$\approx$250,000~km.
The wave speed is estimated to be $> 1050$~km ~s$^{-1}$ for the average oscillation period of $\sim$240 s.
If we assume that the plasma density is $10^{10}$ cm$^{-3}$
({\it 20}),
the implied magnetic field strength is $\sim$50 G for the propagating Alfv\'en wave,
in agreement with measurements and models of active region prominence magnetic fields 
({\it 21, 22}).
The Poynting flux carried by the observed waves is then estimated to be $\rho v^2 V_A\sim 2.0\times 10^6$ erg~s$^{-1}$~cm$^{-2}$
(where $\rho$ is the density, $v$ is the velocity amplitude, and $V_A$ is the Alfv\'en speed),
a lower limit based on the minimum estimates of observed tangential velocities. Given a suitable dissipation mechanism,
this flux is sufficient to heat coronal loops with lengths longer than the estimated oscillation wavelength 
({\it 23}).

In transiting from the photosphere to the chromosphere, the large density decrease with height results in rapid increases in the acoustic and Alfv\'en wave propagation speeds.
This effective discontinuity causes waves with periods longer than a cutoff period determined 
by the thermodynamic conditions and magnetic field strength in the atmosphere to be reflected before reaching the coronal heights of active region prominences.
Since we find a typical oscillation period $\sim$240--250 s, we can infer that the Alfv\'en cut-off period is longer than about 4 minutes for this prominence structure.

The limited field of view of our data prevents us from determining the lengths of the field lines threading the prominence.
However, the frequency spectrum of the observed waves can be used to investigate the length of the field lines. 
Open field lines have a distinct wide spectrum above the Alfv\'en cutoff frequency 
whereas shorter closed loops show multiple discrete resonances as a function of loop length.
The estimated Alfv\'en speed and the mean observed period implies a minimum length of 250,000~km if this is the standing wave in a closed loop system.

Previous observations of waves in the solar corona include Doppler velocity and periodic intensity oscillations in coronal loops as well as
flare-generated transversal displacements of active region loops 
({\it 24-28}).
Those oscillations are examples of magneto-acoustic waves propagating out from photospheric source sites. 
Alfv\'en waves in coronal loops and prominences are also claimed to account for spectroscopic observations of nonthermal line widths in coronal emission lines 
({\it 29, 30}).

\begin{quote}
{\bf References and Notes}

\begin{enumerate}

\item
Y. Lin, O. Engvold, L. Rouppe van der Voort, J. E. Wiik, T. E. Berger,
{\it Sol. Phys.} {\bf 226}, 239-254 (2005).



\item
J. C. Vial, P. Gouttebroze, G. Artzner, P. Lemaire,
{\it Sol. Phys.} {\bf 61}, 39-59 (1979).


\item
J. M. Malherbe, B. Schmieder, E. Ribes, P. Mein,
{\it Astron. Astrophys.} {\bf 119}, 197-206 (1983).

\item
G. Simon, B. Schmieder, P. D\'emoulin, A. I. Poland,
{\it Astron. Astrophys.} {\bf 166}, 319-325 (1986).

\item
B. Schmieder, M. A. Raadu, J. E. Wiik,
{\it Astron. Astrophys.} {\bf 252}, 353-365 (1991).

\item
S. F. Martin,
{\it Sol. Phys.} {\bf 182}, 107-137 (1998).

\item
J. B. Zirker, O. Engvold, S. F. Martin,
{\it Nature} {\bf 396}, 440-441 (1998).

\item
J. Chae, C. Denker, T. J. Spirock, H. Wang, P. R. Goode,
{\it Sol. Phys.} {\bf 195}, 333-346 (2000).

\item
J. T. Karpen, S. K. Antiochos, J. A. Klimchuk,
{\it Astrophys. J.} {\bf 637}, 531-540 (2006).

\item
C. J. Schrijver {\it et al.},
{\it Sol. Phys.} {\bf 187}, 261-302 (1999).

\item
S. Patsourakos, J. C. Vial,
{\it Sol. Phys.} {\bf 208}, 253-281 (2002).

\item
T. Kosugi {\it et al.},
{\it Sol. Phys.} {\bf 243}, 3-17 (2007)

\item
S. Tsuneta {\it et al.},
{\it Sol. Phys.}, http://arxiv.org/abs/0711.1715.






\item
Y. Suematsu, R. Yoshinaga, N. Terao, T. Tsubaki,
{\it Publ. Astron. Soc. Japan.} {\bf 42}, 187-203 (1990).

\item
M. Kuperus, M. A. Raadu,
{\it Astron. Astrophys.} {\bf 31}, 189-193 (1974).

\item
J. L. Leroy, V. Bommier, S. Sahal-Br\'echot,
{\it Astron. Astrophys.} {\bf 131}, 33-44 (1984).

\item
T. Hirayama,
{\it Sol. Phys.} {\bf 100}, 415-434 (1985).

\item
V. M. Nakariakov, E. Verwichte,
{\it Living Rev. Sol. Phys.} {\bf 2}, 3 (2005).

\item
A. J. D\'{\i}az, R. Oliver, J. L. Ballester,
{\it Astrophys. J.} {\bf 580}, 550-565 (2002).

\item
T. Hirayama,
in {\it Coronal and Prominence Plasma}, A. I. Poland, Ed., (NASA Conference Publication No. 2442, 1986), p. 149.

\item
E. Tandberg-Hanssen, J. M. Malville,
{\it Sol. Phys.} {\bf 39}, 107-119 (1974).

\item
E. Wiehr, G. Stellmacher, 
{\it Astron. Astrophys.} {\bf 247}, 379-382 (1991).


\item
G. L. Withbroe, R. W. Noyes, 
{\it Ann. Rev. Astron. Astrophys.} {\bf15}, 363-387 (1977).






\item
I. De Moortel, J. Ireland, A. W. Hood, R. W. Walsh,
{\it Astron. Astrophys.} {\bf 387}, L13-L16 (2002).

\item
C. J. Schrijver, M. J. Aschwanden, A. M. Title,
{\it Sol. Phys.} {\bf 206}, 69-98 (2002).

\item
M. J. Aschwanden, B. De Pontieu, C. J. Schrijver, A. M. Title,
{\it Sol. Phys.} {\bf 206}, 99-132 (2002).


\item
D. Banerjee, R. Erd\'elyi, R. Oliver, E. O'Shea,
{\it Sol. Phys.} {tmp}, 136 (2007).

\item
Y. Lin, O. Engvold, L. H. M. Rouppe van der Voort, M. van Noort,
{\it Sol. Phys.} {\bf tmp}, 71 (2007).

\item
S. Koutchmy, Y. D. \v{Z}ug\v{z}da, V. Loc\v{a}ns,
{\it Astron. Astrophys.} {\bf 120}, 185-191 (1983).

\item
H. Hara, K. Ichimoto,
{\it Astrophys. J.} {\bf 513}, 969-982 (1999).



\item
The authors thank H. Shibahashi, T. Sekii, R. Erd\'elyi, and V. Nakariakov for comments.
\emph{Hinode} is a Japanese mission developed and launched by ISAS/JAXA, with NAOJ as domestic partner and NASA and Science and Technology Facilities Council (STFC) (UK) as international partners.
It is operated by these agencies in co-operation with European Space Agency and Norwegian Space Centre (Norway).
This work was carried out at the NAOJ Hinode science center, which was supported by the Grant-in-Aid for Creative Scientific Research, the Basic Study of
Space Weather Prediction (head investigator, K. S.) from the Ministry of Education, Culture, Sports, Science,
and Technology, Japan,  donation from the Sun Microsystems Incorporated, and NAOJ internal funding.
The National Center for Atmospheric Research is sponsored by the NSF.
T.J.O. is supported by research fellowships from the Japan Society for the Promotion of Science for Young Scientists.
\end{enumerate}
\end{quote}

\section*{Supporting Online Material}

www.sciencemag.org/cgi/content/full/318/5856/1577/DC1

Movie S1

21 May 2007; accepted 06 November 2007

10.1126/science.1145447

\

\begin{figure}[htbp]
   \begin{center}
   \includegraphics{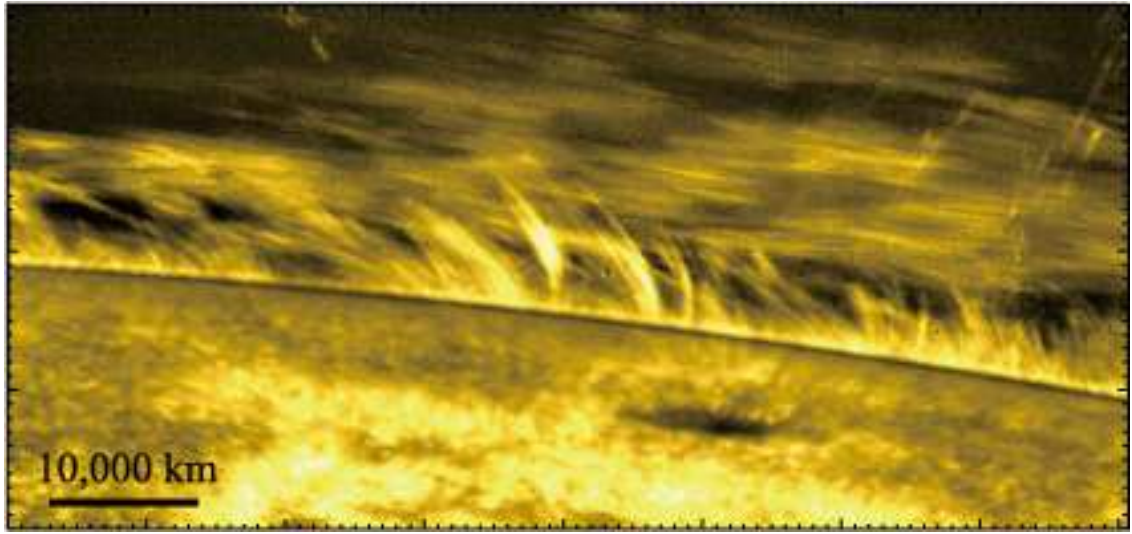}
    \caption{
High-resolution image on the solar limb obtained with SOT aboard \emph{Hinode}.
This observation was performed with a cadence of 15 seconds from 19:33 to 20:44 UT on 9 November 2006.
Tickmarks have a spacing of 1000 km on the Sun.
A radial density filter is applied to show the brighter photosphere and the fainter coronal structures in the same image.
The main sunspot of NOAA AR 10921 as well as the trailing bright plage areas are visible on the disk.
Above the limb, ubiquitous vertical spicules are seen below the horizontal threads of the AR prominence.
The cloudlike prominence structure is located 10,000 to 20,000 km above the visible limb and exhibits a very complex fine structure with predominant horizontal thread-like features.
The intensity of the prominence in Ca II H-line radiation is about 1\% of the on-disk photosphere.
}
   \end{center}
\end{figure}

\

\begin{figure}[htbp]
   \begin{center}
   \includegraphics[width=15cm]{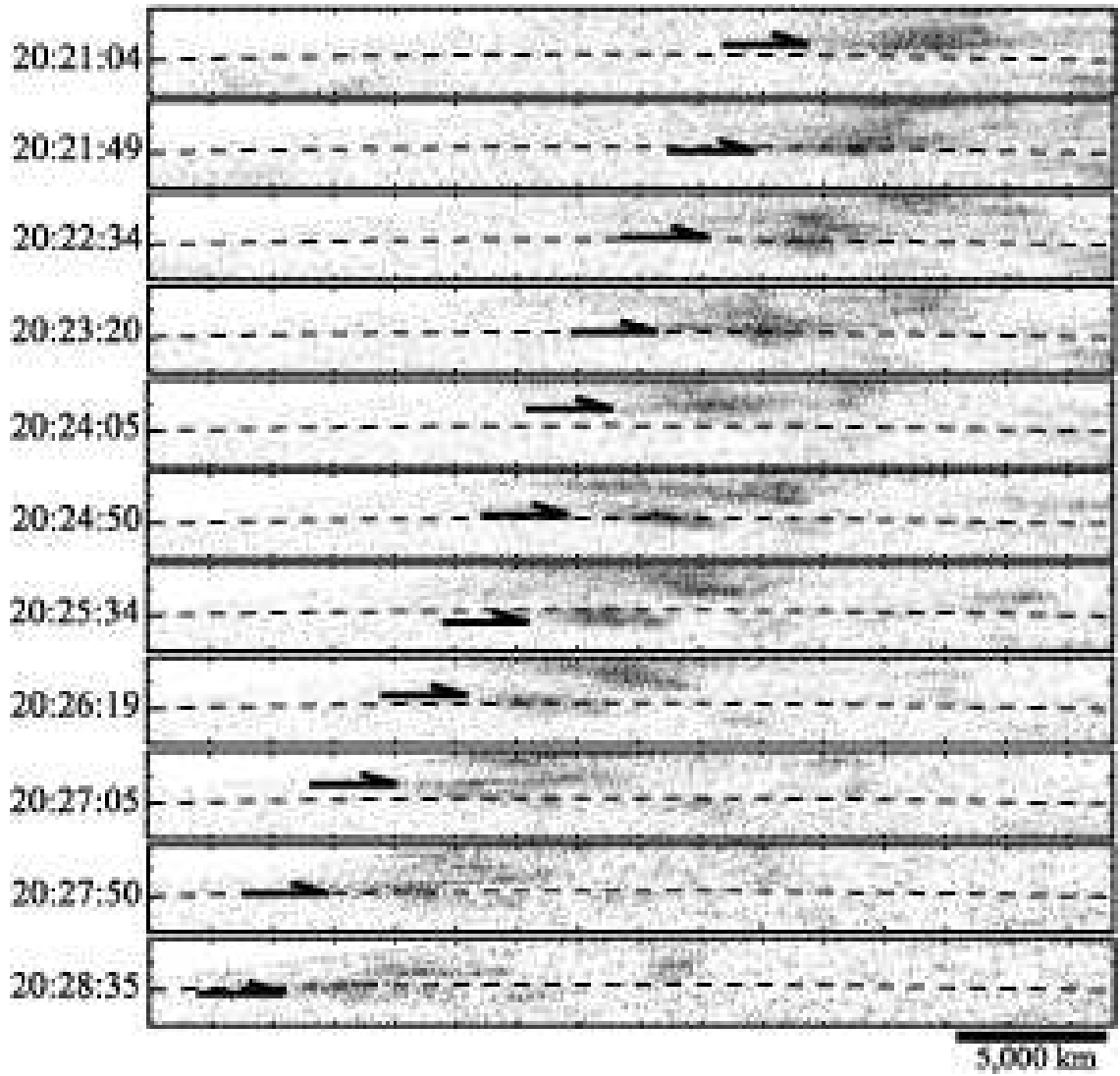}
    \caption{
Example of a vertical oscillation of a single prominence thread.
A small field of view is extracted from the larger field shown in Fig. 1 and shown in negative contrast.
Tickmarks have a spacing of 2000 km on the Sun and UT time for each image is denoted on the left.
The dashed line in each image indicates an approximately constant height above the photosphere.
This oscillating thread has a length of about 3600~km and width of 430~km. Steady flow at 39 km s$^{-1}$ along the thread is evident.
The vertical amplitude of the oscillation is about 900~km with a period of 174~s. The vertical speed is about 10~km~s$^{-1}$. 
}
   \end{center}
\end{figure}

\

\begin{figure}[htbp]
   \begin{center}
   \includegraphics[width=15cm]{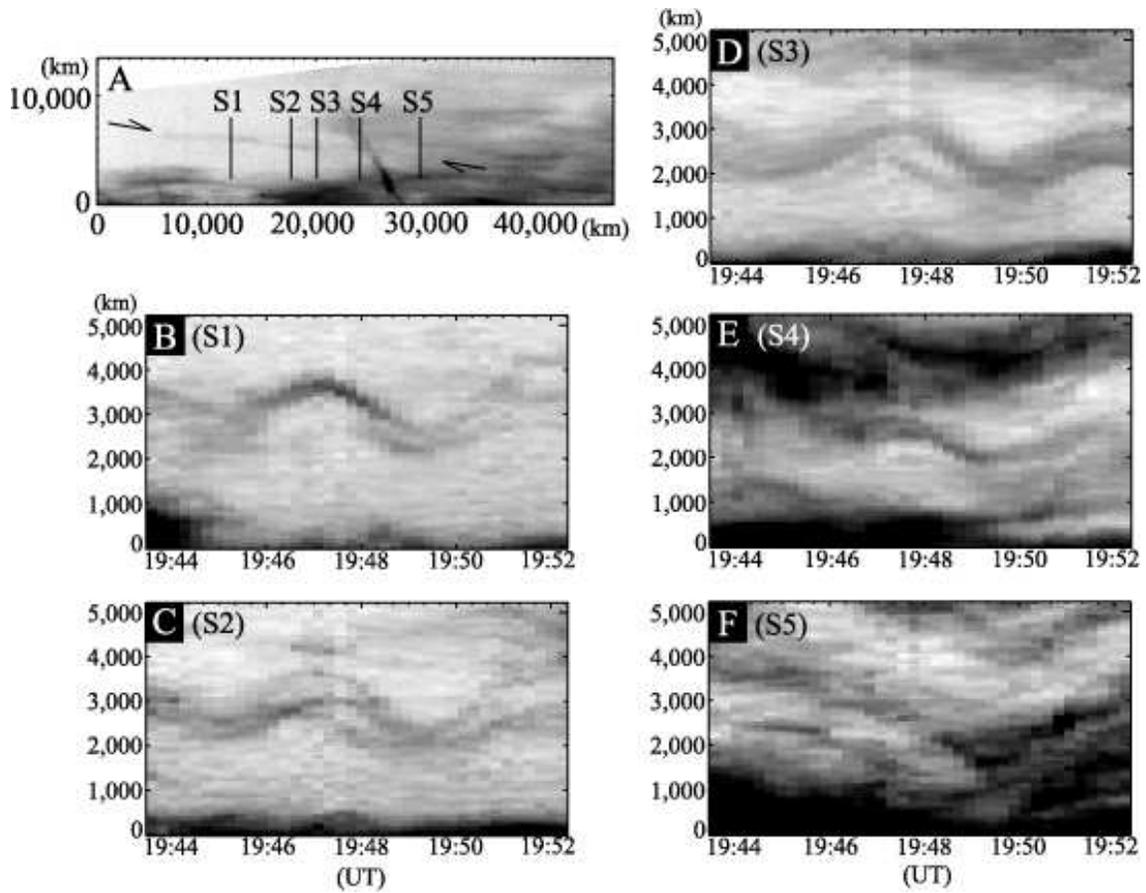}
    \caption{
Example of a prominence thread undergoing synchronous oscillation along its entire length.
({\bf A}), The long thread extending $\sim$16,000 km. Lines S1 to S5 indicate the locations of height versus time plots shown in (B to F).
({\bf B to F}), Height-time plots (shown in negative contrast)  for the locations indicated in A.
Maximum and minimum amplitudes occur at nearly the same time for all locations.
}
   \end{center}
\end{figure}

\begin{table}
\begin{center}
\begin{tabular}{cccccccl}
\\ \hline
 & Length & Width & Horizontal velocity & Vertical oscillation& Vertical oscillation & Height from \\
     & (km)    & (km)  & (km$^{-1}$)                  & period (s)         &  width (km)          & the limb (km) \\
1 & 3600 & 430 & 39 & 174$\pm$25 & 904 & 18,300 \\
2 & 16000 & 660 & 15 & 240$\pm$30 & 1113 & 12,400 \\
3 & 6700 & 580 & 39 & 230$\pm$87 & 909 & 14,700 \\
4 & 2200 & 360 & 46${*}$ & 180$\pm$137 & 435 & 19,000 \\
5 & 3500 & 430 & 45${*}$ & 135$\pm$21 & 408 & 14,300 \\
6 & 1700 & 510 & 25${*}$ & 250$\pm$17 & 1771 & 17,200\\
\hline
\end{tabular}
\end{center}
\caption{Properties of moving threads with vertical oscillations. Asterisks mean averaged velocity.}
\end{table}

\end{document}